\documentclass[conference]{IEEEtran}
\usepackage{cite}
\usepackage{amsmath,amssymb,amsfonts}
\usepackage{algorithmic}
\usepackage{graphicx}
\usepackage{textcomp}
\usepackage{xcolor}
\usepackage{color}

\newcommand{\red}[1]{\textcolor{black}{#1}}
\newcommand{\blue}[1]{\textcolor{black}{#1}}
\newcommand{\yellow}[1]{\textcolor{black}{#1}}
\newcommand{\purple}[1]{\textcolor{black}{#1}}

\def\BibTeX{{\rm B\kern-.05em{\sc i\kern-.025em b}\kern-.08em
    T\kern-.1667em\lower.7ex\hbox{E}\kern-.125emX}}
\begin{document}

\title{Assessment of Subjective and Objective Quality of Live Streaming Sports Videos}

\author{\IEEEauthorblockN{Zaixi~Shang, Joshua P. Ebenezer, Alan C. Bovik}
\IEEEauthorblockA{\textit{Electrical and Computer Engineering}, \\
\textit{The University of Texas at Austin}\\
Austin, Texas \\
\{zxshang, joshuaebenezer\}@utexas.edu, bovik@ece.utexas.edu}
\and
\IEEEauthorblockN{Yongjun Wu, Hai Wei, Sriram Sethuraman}
\IEEEauthorblockA{\textit{Amazon.com, Inc.} \\
Seattle, Washington \\
\{yongjuw, haiwei, sssethur\}@amazon.com}\thanks{J.P. Ebenezer and Z. Shang contributed equally to this work}
}
\maketitle

\begin{abstract}

Video \red{l}ive streaming is gaining prevalence among video streaming service\red{s}, especially for \red{the delivery of popular} sport\red{ing} events. Many objective Video Quality Assessment (VQA) models have been developed to predict the perceptual quality \red{of videos. }Appropriate databases that exemplify the distortions encountered in live streaming videos are important to designing and learning objective VQA models. Towards making progress in this direction, we built a video \red{quality} database \red{specifically} \red{designed} for live streaming VQA \red{research}. The \red{new} video database is called \red{the} Laboratory for Image
and Video Engineering (LIVE) Live stream Database. The LIVE Livestream Database includes 315 videos of 45 contents \red{impaired by} 6 types of distortions. We \red{also} performed a subjective quality \red{study using} the new database\red{, whereby more than} 12,000 human \red{opinions were} gathered from 40 subjects. We \red{demonstrate the usefulness of the new resource by performing} a holistic evaluation of the \red{performance} \purple{of} current state-of-the-art \red{(SOTA)} VQA models. The LIVE Livestream database \red{is being} made publicly available for these purposes at https://live.ece.utexas.edu/research/LIVE\_APV\_Study/apv\_index.
html.

\end{abstract}


\section{Introduction}
There is a variety of factors that can \red{adversely} affect the quality of live streaming videos. For example, bandwidth and stability may affect the received video source quality \red{because of} \purple{variations in} compression\red{, stalls, or scaling.} Compression can \red{cause blocking, banding, motion mismatches, and} local flicker\cite{ni2011flicker}, while scaling \red{can lead to aliasing or interpolation artifacts}\cite{keating1993image}. \red{If} the network connection is unstable or the bitrate \blue{inadequate, then} frame drop\red{s} may \purple{also} occur. If a legacy capture device is used, the source videos might be distorted \red{by} interlacing or judder\red{, especially when there is a rapid motion.} \red{If} the video content is acquired in an interlaced format, \purple{and is then} deinterlaced, the \red{resulting} video \red{may exhibit} combing effects, flicker or noticeable line movements.

\red{The development of} video quality assessment (VQA) models and \red{datasets} has been an \red{ongoing effort for two decades}\cite{seshadrinathan2010study,moorthy2012video,tomar2006converting,bampis2017study,ghadiyaram2017capture,chen2014adaptive,chen2013dynamic,jayaraman2012objective,moorthy2012subjective,chen2019study,tu2020ugc,ChenVMAFC2021,tu2021rapique}. While objective VQA models aim to predict the perceptual quality of videos without the involvement of humans \cite{7115917,9053634}, human subjective quality studies \red{make it possible} to \red{better} understand \red{and model the }specific factors \red{that} contribute to the perceived quality of streaming videos. This \red{data can be used to design or learn} objective  models \purple{that are} consistent with subjective human evaluations \purple{of quality}. 
There have been many efforts to build subjective video quality databases. \red{Among those, the} LIVE VQA Database\cite{seshadrinathan2010study}, the LIVE QoE Database for HTTP-based Video Streaming \cite{chen2014modeling} and the LIVE Mobile Video Quality Database\cite{moorthymobile} consists of various video distortions encountered in video streaming including compression, packet-losses, and video stalls. Similarly, the MCL-V database\cite{lin2015mcl}, and the TUM databases\cite{keimel2010visual}, \red{contain} several synthesized videos \red{with} H.264 compression. 

\red{Yet none of these databases are specifically designed for live streaming distortions}. \red{Among existing datasets, most include fewer} than 20 pristine \red{source} video contents \red{of Standard Definition (SD) or High Definition (HD) resolutions, along with} various distorted versions of them. The distortions \red{in these resources are largely} limited to compression and aliasing, \purple{and the datasets} lack \red{other} live streaming \red{distortions. What is needed is a database of} higher resolution \red{(UHD), high-quality source videos that have been processed to include distortions characteristic of those encountered in live streaming} \purple{scenarios}. 

\red{Towards filling this gap, we have created a new resource that we call the LIVE Livestream Database, which includes} a large number of high motion sports videos, \red{impaired by} the most common distortions that impact the perceptual quality of \red{live streamed} videos. The \red{new} database \red{contains} 315 videos, \red{impaired by six types of common processing} distortions. The LIVE Livestream database consists of Full High Definition (FHD) and Ultra High Definition (UHD) videos of \red{high-motion} sports \red{content captured by professional videographers. Using these videos, we conducted a large human subjective study, whereby we presented the videos} to a large pool of \purple{volunteers} to \red{obtain} Mean Opinion Scores (MOS). \red{To demonstrate the usefulness of the new dataset, we used it} to perform a holistic evaluation of current state-of-the-art VQA models, to \red{compare their} performance and \red{to gain insights into} potential future \red{live streaming} VQA \red{problems}. 

The rest of the paper is organized as follows: In Section~\ref{details}, we explain the details \red{of} the \red{construction of} new database and the \red{protocol of the} human study. Section \ref{compare} compares the performances of various \blue{state of the art (SOTA)} VQA models on \red{the} new database. Finally, Section \ref{conclusion} concludes the \red{paper with thoughts regarding future efforts}.

\section{SUBJECTIVE STUDY}\label{details}

\subsection{Source Sequences}
\begin{figure*}[!t]
\centering
{\includegraphics[width=0.18\linewidth,height=52pt]{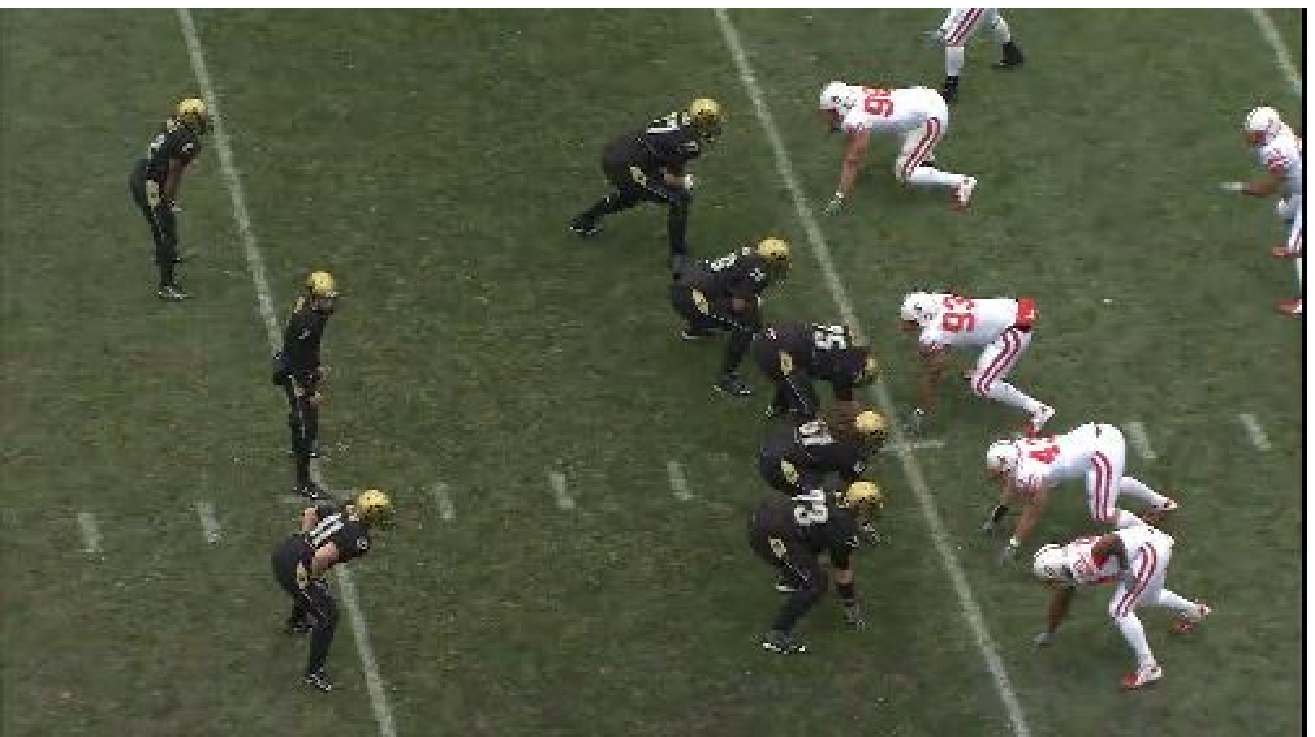}}
{\includegraphics[width=0.18\linewidth,height=52pt]{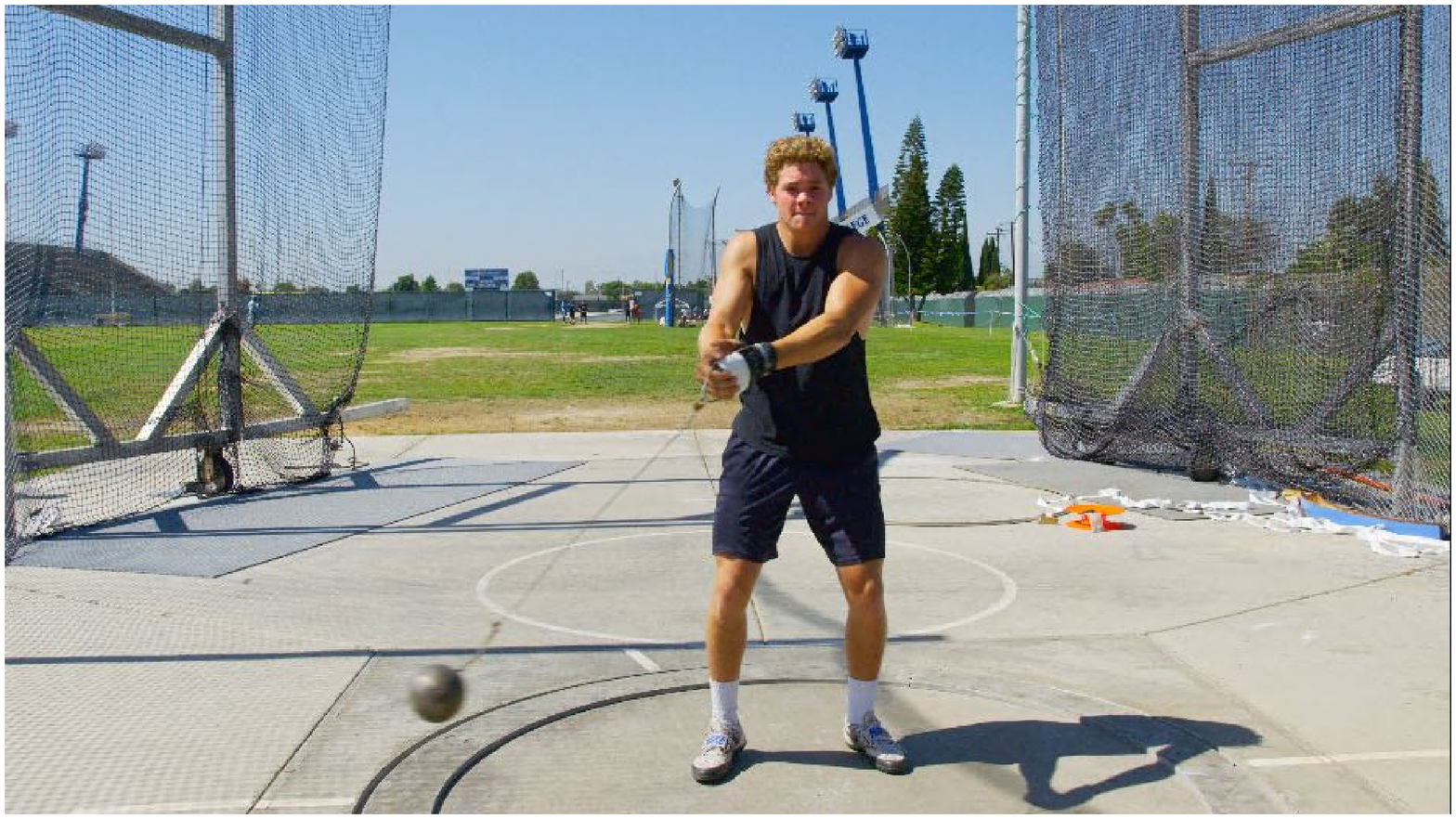}}
{\includegraphics[width=0.18\linewidth,height=52pt]{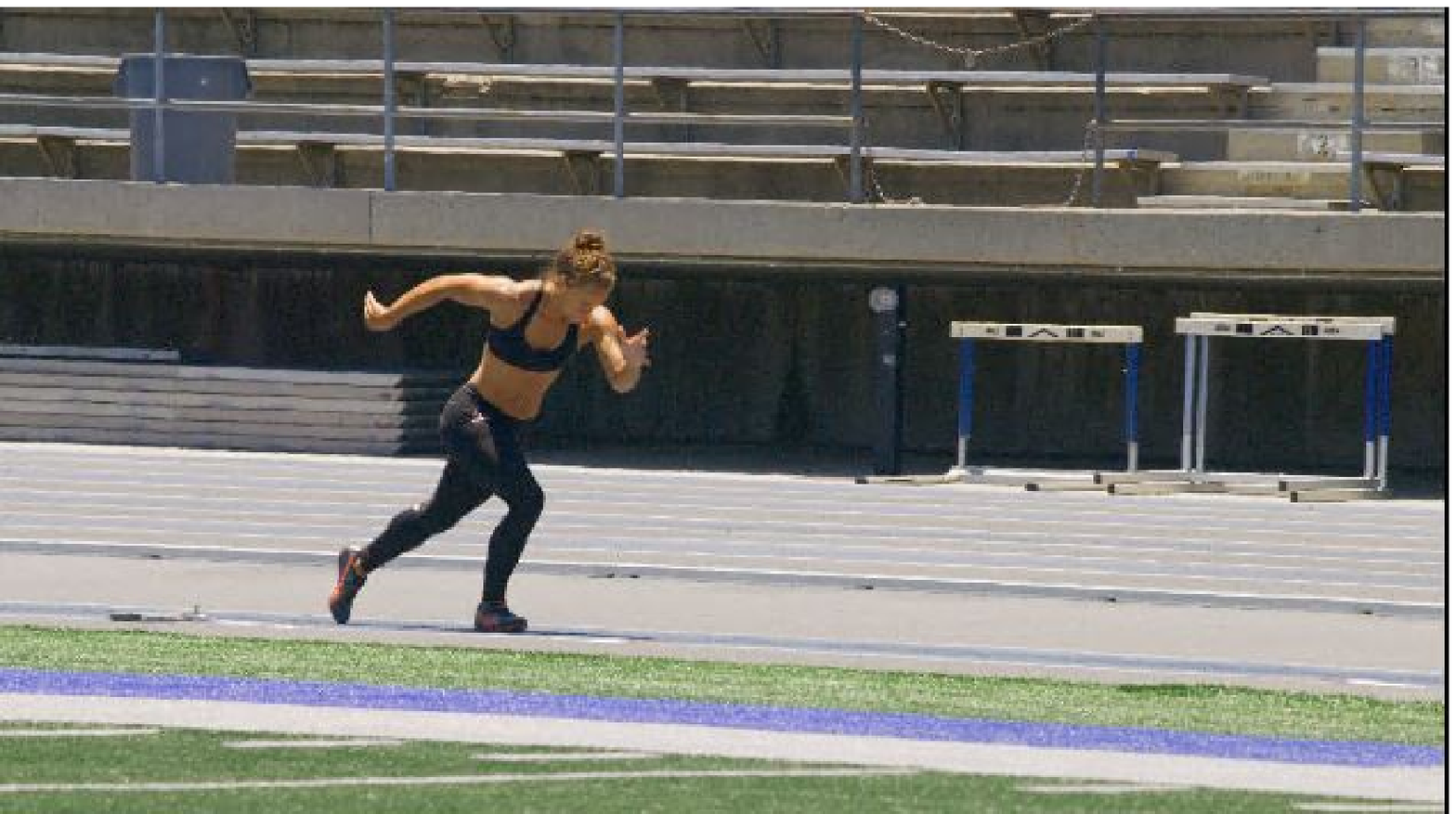}}
{\includegraphics[width=0.18\linewidth,height=52pt]{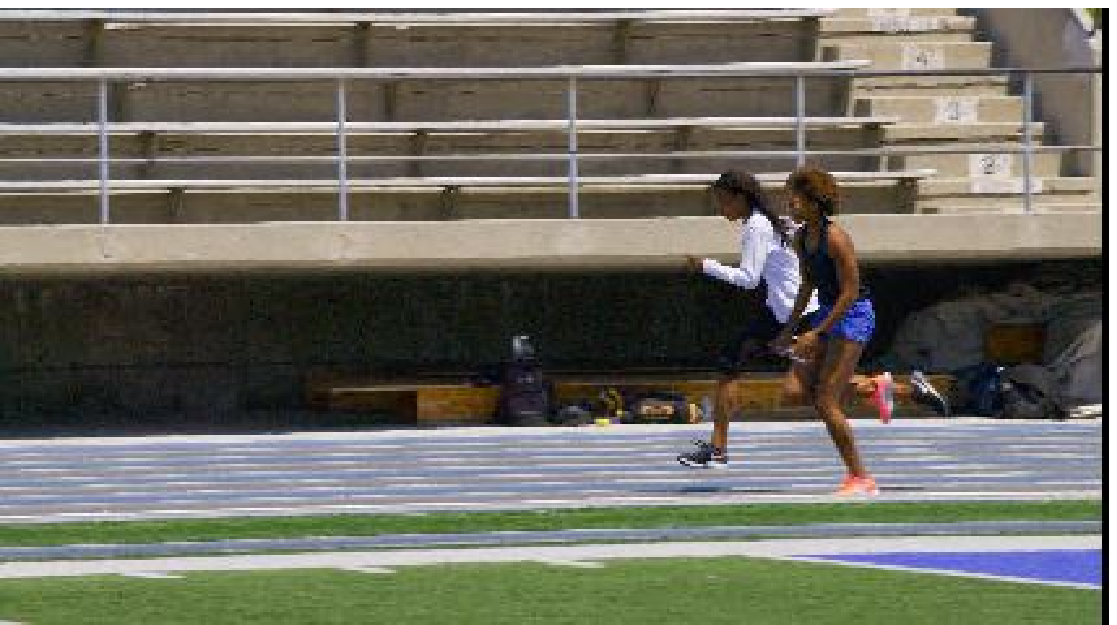}}
{\includegraphics[width=0.18\linewidth,height=52pt]{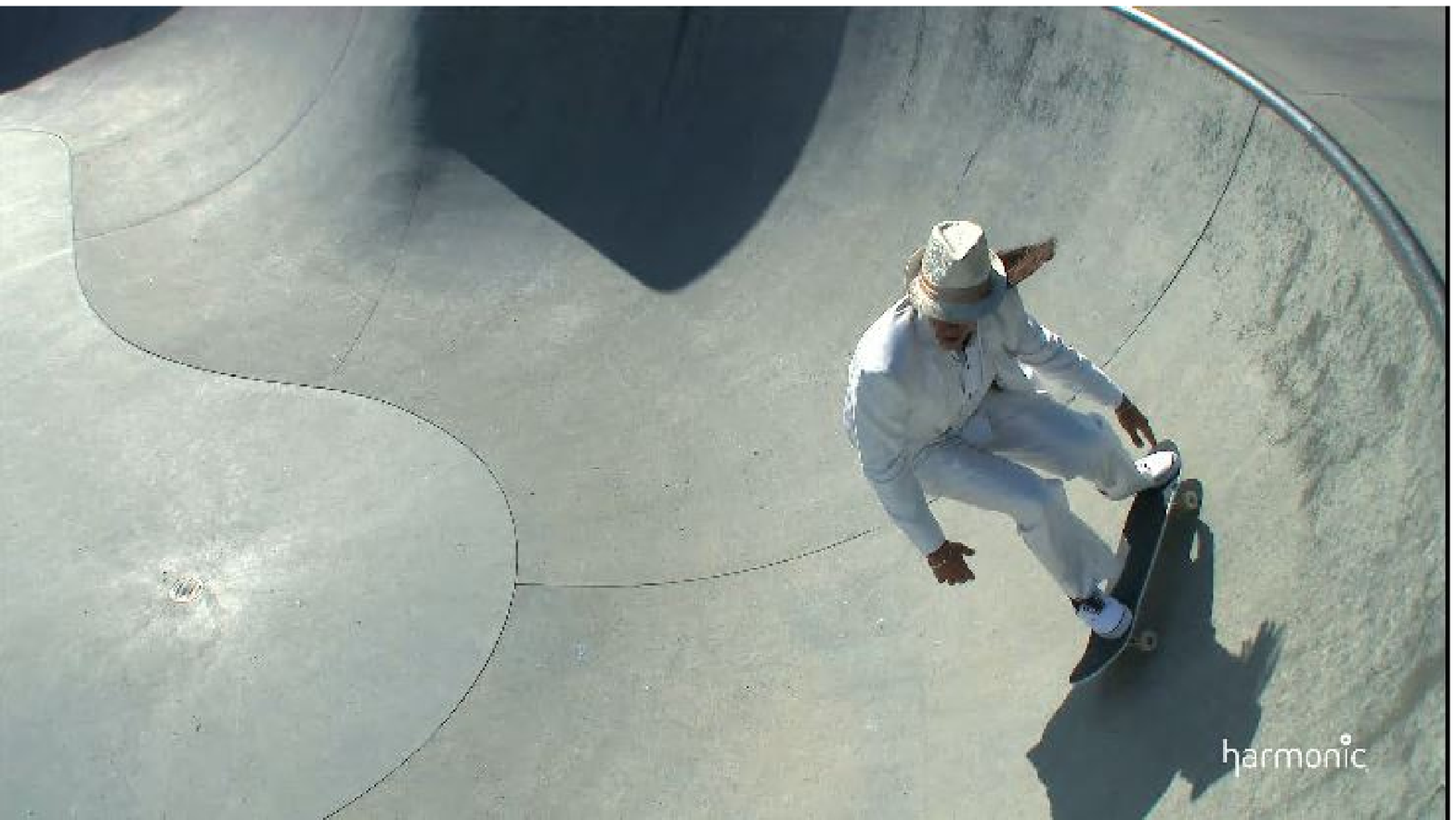}}
\caption{Sample screen shots from the gathered video sequences.}
\label{example}
\end{figure*}

We collected 33 uncompressed, high-quality, freely available online videos from multiple sources, including \red{from} Tampere University\cite{mercat2020uvg}, the MCML Group\cite{cheon2017subjective}, the Netflix Public Dataset\cite{li2016toward}, \red{the} VQEG HD3 Dataset\cite{video2000final}, \red{the} Consumer Digital Video Library (CDVL)\cite{yodel2011consumer}, and \red{the} SJTU Media Lab\cite{song2013sjtu}. All \red{of the selected} videos were \red{captured} with professional, high-end \red{camera} equipment and are distortion-free. The original pristine videos all have resolutions of 1920x1080 or 3840x2160 pixels, and were progressively scanned in YUV 4:2:0 format with audio components removed. The videos have frame rates ranging from 25 fps to 30 fps. The \red{video} contents \red{include} 15 different types of sports, including running, football, and soccer, and one video of the audience in a stadium, as exemplified in Fig. \ref{example}.

The \purple{original} 33 videos that we collected \red{are} of durations ranging from 5s to 26s, and we manually cropped the original videos into shorter clips  \red{of about} 7 seconds. 45 video clips \red{were} created \purple{from the 33 originals}, of which 22 clips are of resolution 1920x1080 and 23 clips are of resolution 3840x2160.

\subsection{Synthetic Distortions}
Six distorted video sequences were created from each of the pristine sequences, using six different distortion processes. \red{These included} H.264 compression, aliasing, judder, flicker, frame drops, and interlacing. \red{When applying different} levels of \red{each} distortion type, we \red{sequenced through the} reference sequences such that each \red{would} have only \red{a single severity level of each} distortion type. For example, \red{four levels of } H.264 compression, corresponding to different constant rate factors (CRF) \red{were defined.} The first reference \red{video could only} be compressed \red{using} the first CRF level, the second reference was only compressed \red{using} the  second \red{CRF} level, and so on. The fifth \red{source} video \red{then had the} first level \red{of distortion applied.} \red{However,} to ensure \red{that there would be } no content-related quality bias, the \red{first} video \red{in} the quality level cycle \red{was also sequenced} \purple{as subsequent distortions were applied.} \red{In this way}, each \purple{of the 45 clips taken from the original 45 pristine source videos} \red{has 6 associated distorted versions of it, yielding 315 videos. }
\subsubsection{H.264 Compression (c)}

H.264 \red{remains} the most \red{widely-}accepted and used video compression standard. We \red{fixed} four levels of H.264 compression using the criteria described earlier\red{, by varying the CRF values.} \red{CRF level 1 videos} have perceptual qualities similiar to the reference videos, \red{while} those of the worst (CRF level 4) \red{videos exhibit} obvious \red{compression artifacts.} All of the compressed videos \red{were} generated using FFmpeg.
\subsubsection{Aliasing}
Aliasing was simulated by first downscaling \red{each} video, then upscaling it back to its original dimensions. The downscaling was performed by \red{spatially} downsampling the video to half the original size without the use of an anti-aliasing filter, while the upscaling was performed using a Lanczos filter. 
\subsubsection{Judder}
\begin{figure}[!t]
\centering
{\includegraphics[width=0.13\linewidth,height=145pt]{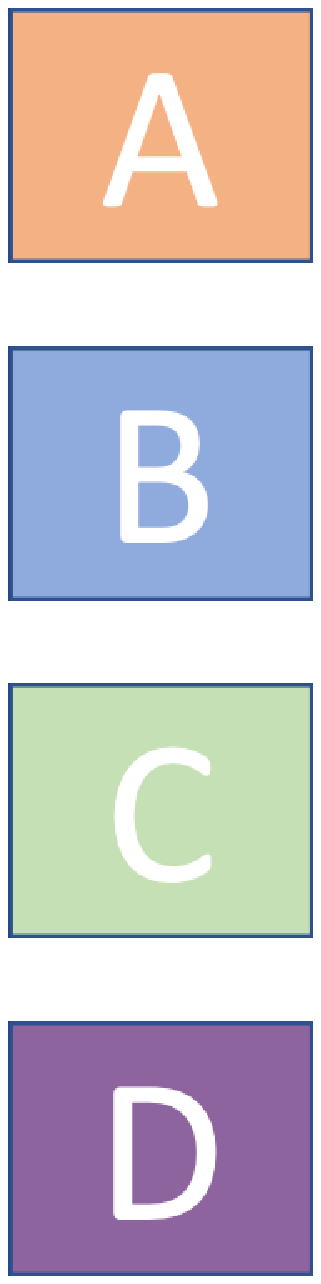}%
\label{fig:side:a}}
\hfil
{\includegraphics[width=0.24\linewidth,height=5cm,keepaspectratio]{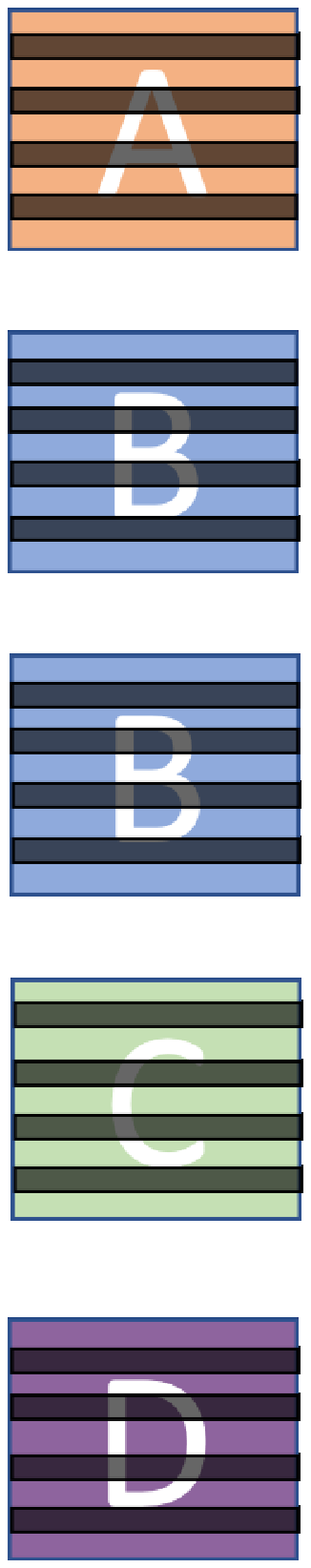}%
\label{fig:side:a}}
\hfil
{\includegraphics[width=0.24\linewidth,height=5cm,keepaspectratio]{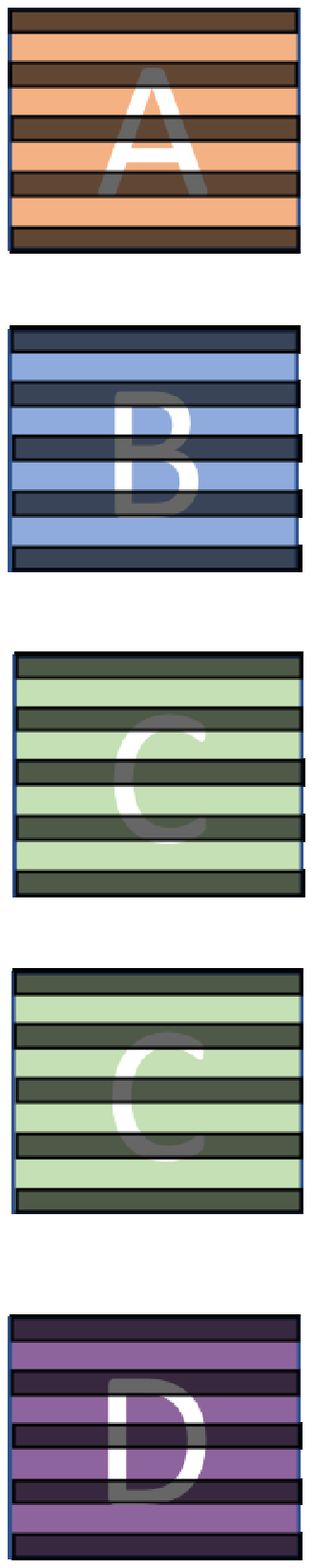}%
\label{fig:side:a}}
\hfil
{\includegraphics[width=0.24\linewidth,height=5cm,keepaspectratio]{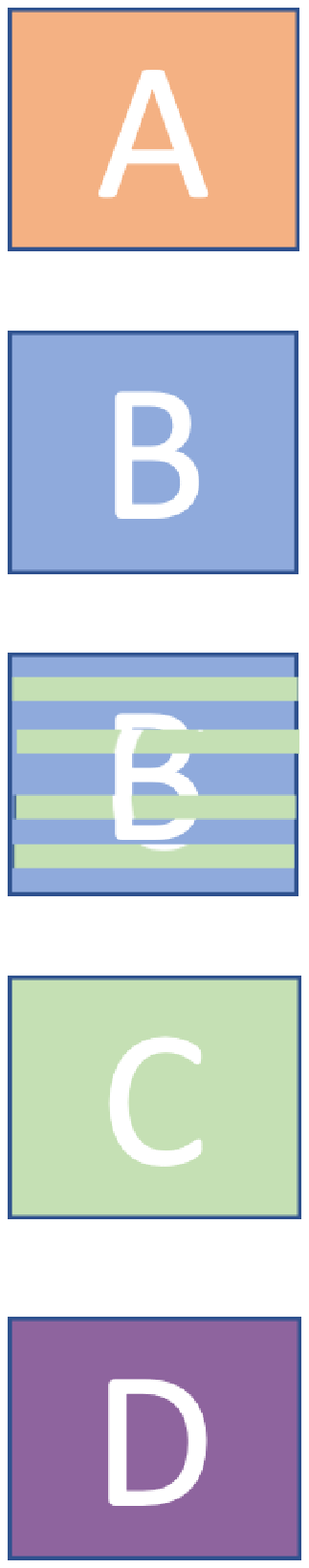}%
\label{fig:side:a}}

\caption{Simulation of motion judder \red{from} 3:2 pulldown. (a) Original frames at 23.94 fps. (b) Odd video fields. (c) Even video fields (d) Resulting frames formed by interlacing odd and even fields at 29.97 fps}\label{fig:judder}
\end{figure}

Motion judder is an artifact that is introduced when scenes shot at 23.94 fps are converted to 29.97 fps by a process called 3:2 pulldown. The ratio of these frame rates is 4:5: for every 4 input frames, 5 output frames were created by temporally downsampling the video to 23.94 fps using FFMPEG, then converting the frame rate to 29.97 by 3:2 pulldown. The odd video field of every 2\textsuperscript{nd} frame, and the even video field of every 3\textsuperscript{rd} frame \red{of} each group of 4 frames \red{were} combined to form an additional frame, for each group of 4 frames. \red{This} process is shown in Fig. \ref{fig:judder}. 

\begin{figure*}[!t]
\centering
{\includegraphics[width=0.3\linewidth]{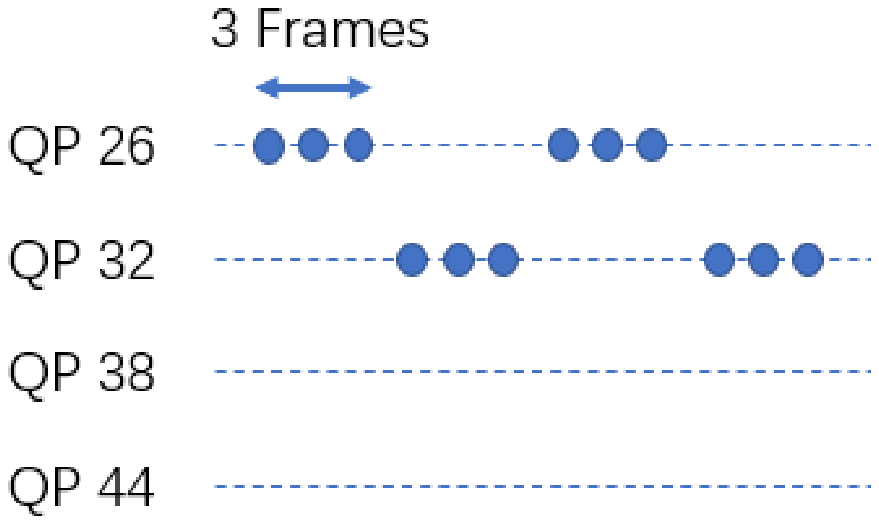}}
{\includegraphics[width=0.3\linewidth]{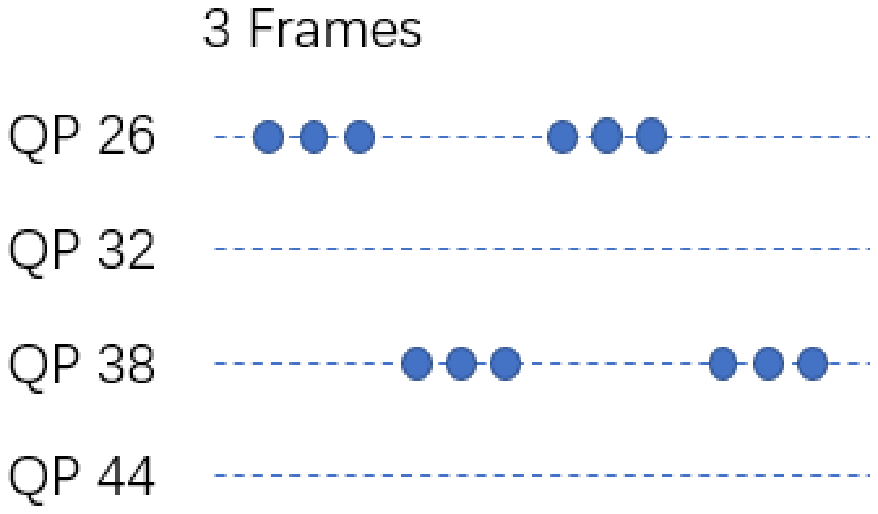}}
{\includegraphics[width=0.3\linewidth]{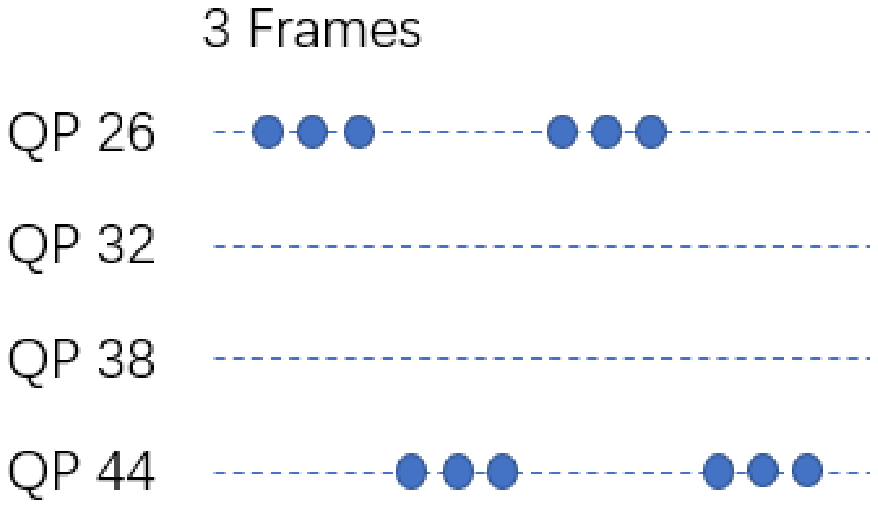}}
\caption{Three levels of flicker synthesis. Video flicker is generated by altering the QP levels on the videos. Three combinations of QP values are used to generate different flicker levels.}
\label{flicker}
\end{figure*}

\subsubsection{Flicker}
We simulated flicker distortion \red{from compression} by alternating the \red{H.264} quantization parameter (QP) on the video. Three pairs of QPs were chosen to form three \red{flicker} distortion levels:  QP26 and QP32, QP26 and QP38, and QP26 and QP44. The flicker rate, which is the number of QP alternations per second, \red{was} kept a constant roughly 5 Hz i.e. by alternating the QP \red{every} 3 frames. \red{This process is depicted} in Fig. \ref{flicker}.

\subsubsection{Frame Drops}
We simulated video frame loss\red{es} that \red{occur} when \red{a} source video is transmitted over a channel, such as
a wireless network. We simulated frame drop cluster\red{s} \yellow{of adjacent frames} to account for 10\%-30\% of a group of pictures (GOP). Three levels of frame drop \red{densities were} chosen: 3, 6 and 9 frames \red{per} cluster, \red{yielding} a slight to severe \red{impact on} the perceptual \purple{qualities} \red{of} the videos.
\subsubsection{Interlacing (i)}
\red{On} each frame \red{of the video,} the even and odd lines were separated to form two fields, field A and field B. Field B from \red{each} current frame and field A from \red{each} next frame \red{were then} combined to create interlaced frames. \red{In the presence of motion}, combing effects become evident. \red{Since} interlaced video fields \red{are} captured at different moments in time, interlaced frames \red{often} exhibit motion \red{combing} artifacts, \red{when} objects move \red{quickly} enough to be at different positions \red{in} each field.

\subsection{Subjective Testing Design}
\purple{In the human study, a} single-stimulus (SS) method was employed, as described in the ITU-R BT 500.13 recommendation\cite{union2012methodology}. The subjects used a rating bar to \red{record their subjective opinion scores.} After displaying each of the test \red{videos}, a continuous \red{rating bar was} displayed on the screen with a randomly \red{placed} cursor. The quality bar \red{was marked with} labels ``Bad,” ``Poor,” ``Fair,” ``Good,” and ``Excellent” to facilitate the subjects in making decisions. The subjects use a Palette gear console to provide subjective scores. Video rating scores \red{were} given after watching \red{each} video \red{on an (invisible) scale} ranging from 0 to 100, where 0 indicates the worst quality and 100 indicates the best quality.

\subsection{Subjective Testing Environment and Display}
The human study was carried out in \red{the} LIVE \red{Subjective study room} at The University of Texas at Austin. The Lab \red{was arranged} to simulate a living room environment. The windows were covered, and background distractions were removed. A Samsung UN65RU7100FXZA Flat 65-Inch 4K UHD TV \red{was} used to display \red{all of} the videos. The viewing distance was about 2H, where H is the height of the TV so that the subjects \red{could} comfortably \red{view} the videos and \red{assess} the video distortions.

Since the TV \red{is able to} upscale 1080p content \red{using} an unknown algorithm, all \red{of} the 1080p videos \red{were} \purple{instead} upscaled \red{using the} Lanczos resizing \red{function} to avoid any unpredictable effects. \red{The} 1080p videos \red{were upscaled to 4K, after the} distortions were applied. To ensure perfect playback, all \red{of} the \red{videos} were stored as raw YUV 4:2:0 files. \red{The powerful} Venueplayer application developed by VideoClarity was used to guarantee smooth playback of the 4K videos. 

\subsection{Human Subjects and Score Processing}
\red{A total of} 40 \red{human subjects were} recruited \red{from the} student \red{population at The} University of Texas at Austin. Two of the subjects finished \purple{only} one of the two sessions, and the rest 38 human subjects finished both sessions. 154 videos were rated by 40 subjects\red{,} while 161 videos were \red{rated} by 38 subjects. The subject pool was inexperienced with video quality assessment and video distortions. 

\begin{figure}[!t]
\includegraphics[width=0.95\linewidth]{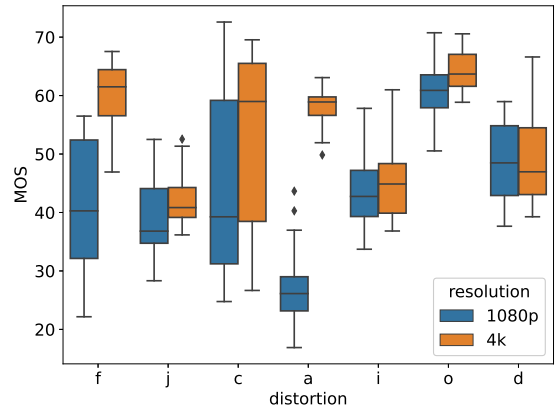}

\caption{Box plot comparing MOS \red{against distortion type for both considered} video resolutions. The labels on \red{the horizontal} axis represent: f: flicker; j: judder; c: compression; a: aliasing; i: interlacing; d: frame drop and o: original (reference videos).}
\label{fig:resolution}
\end{figure}

Subjective Mean Opinion Scores (MOS) \red{were} computed as described in\cite{yu2019predicting}. 
A box plot of the calculated MOS score is shown in Fig. \ref{fig:resolution}. The distorted video classes \red{exhibit} different \red{distributions}, since they \red{reflect} different \red{types and} levels of distortion. \red{The purely temporal distortions--}interlacing, judder, and frame drops--\red{yielded} similiar ranges of MOS for 1080p and 4K videos. \red{However,} aliasing \red{resulted in} very different MOS \red{ranges, likely because of the additional} upscaling \red{of} 1080p videos when displayed on \red{the} 4K TV.

\begin{table*}[!t]
\renewcommand\arraystretch{1.3}
\centering
\caption{SROCC of the Compared NR VQA Models. The Scores of the Top Performing Algorithm Are Boldfaced}\label{tab:sroccall}
\begin{tabular}{|c|c|c|c|c|c|c|c|}
\hline
ALGORITHM&\textsc{Overall}&COMPRESSION&ALIASING&JUDDER & FLICKER&FRAME DROP&INTERLACING\\

\hline
NIQE & 0.3232 &0.3175  & 0.3060  & 0.2863  & 0.3332  &0.2842  & 0.2780 \\
\hline
BRISQUE &0.6381& 0.5748  & 0.7564   & 0.8235  & \textbf{0.6574}   & 0.2569  &  0.8689 \\
\hline
VIIDEO & 0.0044 & 0.0053  & 0.0073  & 0.0055  & 0.0013  & 0.0024  & 0.0064  \\
\hline
CORNIA &0.6778 & \textbf{0.6873}  &\textbf{ 0.7853 } & 0.8390   & 0.5861 & 0.2776  &\textbf{ 0.8864 } \\
\hline
HIGRADE &0.6916 & 0.5748  & 0.6965  & 0.7729  & 0.6295  & 0.6057  & 0.8266  \\
\hline
V-BLIINDS & 0.7330 & 0.6450  & 0.7606  &\textbf{ 0.8679}  & 0.6182  & 0.7131  & 0.8060  \\
\hline
TLVQM & 0.7503 & 0.5614  & 0.7420  & 0.8328  & 0.6202  & \textbf{0.8555} &  0.8173  \\
\hline
\yellow{ChipQA-0} & \textbf{0.7513} & 0.6594  & 0.7791  & 0.8513  & 0.6491  & 0.6780 &  0.8534  \\
\hline
\end{tabular}
\end{table*}

\begin{table*}[!t]
\renewcommand\arraystretch{1.3}
\centering
\caption{PLCC of the Compared NR VQA Models. The Scores of the Top Performing Algorithm Are \red{Boldfaced}.}\label{tab:plccall}
\begin{tabular}{|c|c|c|c|c|c|c|c|}
\hline
ALGORITHM&OVERALL&COMPRESSION&ALIASING&JUDDER & FLICKER&FRAME DROP&INTERLACING\\
\hline
NIQE& 0.4962 &0.4983 & 0.4820 & 0.3860  & 0.2848  & 0.2849  & 0.2850  \\
\hline
BRISQUE&0.6698& 0.7345  & 0.9321  & 0.8726  & 0.7268  & 0.3902   & 0.9118  \\
\hline
VIIDEO & 0.1271 & 0.1222  & 0.1222  & 0.1235  & 0.1247  & 0.1256  & 0.1259  \\
\hline
CORNIA &0.7257 & \textbf{0.8243 } & 0.9472  & 0.8642  &0.5871 &0.2138   &\textbf{  0.9216  }\\
\hline
HIGRADE &0.6990 &0.6913   & 0.9311  & 0.8014  &0.6402   & 0.6025   & 0.8699   \\
\hline
V-BLIINDS&0.7477 & 0.8055  & 0.9202  &  \textbf{0.9200 }& 0.7086  & 0.7443   &  0.8873 \\
\hline
TLVQM &0.7513& 0.6788  & 0.9273   & 0.8914  & \textbf{0.7724}  &\textbf{ 0.8738 }  & 0.8358   \\
\hline
\yellow{ChipQA-0} &\textbf{0.7565} & 0.7783  &\textbf{ 0.9490}   & 0.9071  & 0.6609  &0.6945 & 0.9075   \\
\hline
\end{tabular}
\end{table*}

\section{Objective VQA Model Comparison}\label{compare}
We evaluated several publicly available objective VQA algorithms on the LIVE Livestream Database to demonstrate the usefulness of the new resource. The performances of the objective VQA algorithms \red{were} evaluated \red{using} the Spearman’s Rank Order Correlation Coefficient (SROCC) and the Pearson Linear Correlation Coefficient (PLCC).

\subsection{Performance of VQA Models}
The NR VQA algorithms that \red{were} tested include NIQE\cite{mittal2012making}, BRISQUE\cite{mittal2012no}, HIGRADE\cite{kundu2017no}, CORNIA\cite{ye2012unsupervised}, TLVQM\cite{korhonen2019two}, VIIDEO\cite{mittal2015completely}, V-BLIINDS\cite{saad2012blind}, and ChipQA-0\cite{ebenezer2020noreference}. BRISQUE, HIGRADE, CORNIA, TLVQM, V-BLIINDS, and \yellow{ChipQA-0} are supervised learning algorithms \red{that use} a support vector regressor (SVR) to learn mappings from 'quality-aware' features to mean opinion scores. These algorithms were tested on 1000 random train-test splits. \red{On each split,} 80\% of the data was used for training, and 20\% for testing. \red{Care was taken to ensure that} no content \red{could} appear in both \red{the training} and testing set, or \red{the training and} validation set. NIQE, BRISQUE and HIGRADE are image quality assessment (IAQ) algorithms, so they \red{were} used to extract features frame by frame, followed by temporal \red{average} pooling.

For the unsupervised methods (NIQE and VIIDEO), the scores $s$ \red{were} passed through a nonlinear logistic regression \red{process} before the PLCC, \red{as} described in \cite{yu2019predicting}. The performances of the compared VQA models on the entire database, as well as for each synthetic distortion, are shown in Tables \ref{tab:sroccall} and \ref{tab:plccall}, where the best performing \red{model on} each \red{distortion} category \red{is boldfaced.} The results for \red{each} specific distortion \red{were} acquired by training the SVR on the reference sequences and the \red{specific} distorted sequences. 

\subsection{Discussion of Results}

From Tables \ref{tab:sroccall}, and \ref{tab:plccall}, \red{it may be observed} that \yellow{V-BLIINDS} performed the best among the compared \red{NR VQA} algorithms, \yellow{while TLVQM and ChipQA-0 also achieved relatively higher  correlations against the human judgments.} \red{These learning-based models use simple measurements of motion, which is highly relevant on these kinds of videos.} CORNIA \red{yielded top} performances \red{on} compression, aliasing, and interlacing, \red{all of which present strong spatial aspects of distortion.} However, the overall \red{performance of} CORNIA \red{was} lower than that of V-BLIINDS, TLVQM, and ChipQA-0, due to \red{the lack of} temporal information, e.g., when processing frame drops, TLVQM \red{was able to} effectively capture \red{the perceptual effects of} frame drops, but its performance drops \red{on the} compression and flicker distortions. 

\section{Conclusion}\label{conclusion}
We created a large scale \red{video quality} database \red{targeting high-motion, live streaming scenarios.} The new resource includes 45 different contents and 6 different distortion types. \red{The new database can be used to create, test, and compare both VQA models. We are making the new LIVE Livestream database publicly } available. Future steps include developing new VQA models using the proposed database.

\section*{Acknowledgment}

Z. Shang and J.P. Ebenezer are co-first authors of this work. The authors acknowledge the Texas Advanced Computing Center (TACC) at The University of Texas at Austin for providing {HPC, visualization, database, \red{and} grid} resources that have contributed to the research results reported \red{in} this paper. URL: http://www.tacc.utexas.edu.

\bibliographystyle{IEEEtran}
\bibliography{IEEEabrv,refs}

\begin{thebibliography}{10}
\providecommand{\url}[1]{#1}
\csname url@samestyle\endcsname
\providecommand{\newblock}{\relax}
\providecommand{\bibinfo}[2]{#2}
\providecommand{\BIBentrySTDinterwordspacing}{\spaceskip=0pt\relax}
\providecommand{\BIBentryALTinterwordstretchfactor}{4}
\providecommand{\BIBentryALTinterwordspacing}{\spaceskip=\fontdimen2\font plus
\BIBentryALTinterwordstretchfactor\fontdimen3\font minus
  \fontdimen4\font\relax}
\providecommand{\BIBforeignlanguage}[2]{{%
\expandafter\ifx\csname l@#1\endcsname\relax
\typeout{** WARNING: IEEEtran.bst: No hyphenation pattern has been}%
\typeout{** loaded for the language `#1'. Using the pattern for}%
\typeout{** the default language instead.}%
\else
\language=\csname l@#1\endcsname
\fi
#2}}
\providecommand{\BIBdecl}{\relax}
\BIBdecl

\bibitem{ni2011flicker}
P.~Ni, R.~Eg, A.~Eichhorn, C.~Griwodz, and P.~Halvorsen, ``Flicker effects in
  adaptive video streaming to handheld devices,'' in \emph{19th ACM Int. Conf.
  on Multimedia}, 2011, pp. 463--472.

\bibitem{keating1993image}
S.~M. Keating, ``Image signal process. with digital filtering to minimize
  aliasing caused by image manipulation,'' U.S. Patent 5\,206\,919, Apr. 27,
  1993.

\bibitem{seshadrinathan2010study}
K.~Seshadrinathan, R.~Soundararajan, A.~C. Bovik, and L.~K. Cormack, ``Study of
  subjective and objective quality assessment of video,'' \emph{IEEE Trans.
  Image Process.}, vol.~19, no.~6, pp. 1427--1441, 2010.

\bibitem{moorthy2012video}
A.~K. Moorthy, L.~K. Choi, A.~C. Bovik, and G.~De~Veciana, ``Video quality
  assessment on mobile devices: Subjective, behavioral and objective studies,''
  \emph{IEEE J. Sel. Topics Signal Process.}, vol.~6, no.~6, pp. 652--671,
  2012.

\bibitem{tomar2006converting}
S.~Tomar, ``Converting video formats with ffmpeg,'' \emph{Linux J.}, vol. 2006,
  no. 146, p.~10, 2006.

\bibitem{bampis2017study}
C.~G. Bampis, Z.~Li, A.~K. Moorthy, I.~Katsavounidis, A.~Aaron, and A.~C.
  Bovik, ``Study of temporal effects on subjective video quality of
  experience,'' \emph{IEEE Trans. Image Process.}, vol.~26, no.~11, pp.
  5217--5231, 2017.

\bibitem{ghadiyaram2017capture}
D.~Ghadiyaram, J.~Pan, A.~C. Bovik, A.~K. Moorthy, P.~Panda, and K.-C. Yang,
  ``In-capture mobile video distortions: A study of subjective behavior and
  objective algorithms,'' \emph{IEEE Trans. Circuits Syst. Video Technol},
  vol.~28, no.~9, pp. 2061--2077, 2017.

\bibitem{chen2014adaptive}
C.~Chen, X.~Zhu, G.~de~Veciana, A.~C. Bovik, and R.~W. Heath, ``Adaptive video
  transmission with subjective quality constraints,'' in \emph{IEEE ICIP},
  2014, pp. 2477--2481.

\bibitem{chen2013dynamic}
C.~Chen, L.~K. Choi, G.~de~Veciana, C.~Caramanis, R.~W. Heath, and A.~C. Bovik,
  ``A dynamic system model of time-varying subjective quality of video streams
  over http,'' presented IEEE ICASSP, 2013, pp. 3602--3606.

\bibitem{jayaraman2012objective}
D.~Jayaraman, A.~Mittal, A.~K. Moorthy, and A.~C. Bovik, ``Objective quality
  assessment of multiply distorted images,'' in \emph{Conf. Rec. 46th
  ASILOMAR}, 2012, pp. 1693--1697.

\bibitem{moorthy2012subjective}
A.~K. {Moorthy}, L.~K. {Choi}, A.~C. {Bovik}, and G.~{de Veciana}, ``Video
  quality assessment on mobile devices: Subjective, behavioral and objective
  studies,'' \emph{IEEE J. Sel. Topics Signal Process.}, vol.~6, no.~6, pp.
  652--671, 2012.

\bibitem{chen2019study}
M.~Chen, Y.~Jin, T.~Goodall, X.~Yu, and A.~C. Bovik, ``Study of 3d virtual
  reality picture quality,'' \emph{IEEE Journal of Selected Topics in Signal
  Processing}, vol.~14, no.~1, pp. 89--102, 2019.

\bibitem{tu2020ugc}
Z.~Tu, Y.~Wang, N.~Birkbeck, B.~Adsumilli, and A.~C. Bovik, ``{UGC-VQA}:
  Benchmarking blind video quality assessment for user generated content,''
  \emph{arXiv preprint arXiv:2005.14354}, 2020.

\bibitem{ChenVMAFC2021}
L.-H. Chen, C.~G. Bampis, Z.~Li, J.~Sole, and A.~C. Bovik, ``Perceptual video
  quality prediction emphasizing chroma distortions,'' \emph{{IEEE} Trans.
  Image Process.}, vol.~30, pp. 1408--1422, 2021.

\bibitem{tu2021rapique}
Z.~Tu, X.~Yu, Y.~Wang, N.~Birkbeck, B.~Adsumilli, and A.~C. Bovik, ``Rapique:
  Rapid and accurate video quality prediction of user generated content,''
  \emph{arXiv preprint arXiv:2101.10955}, 2021.

\bibitem{7115917}
S.-C. Pei and L.-H. Chen, ``Image quality assessment using human visual dog
  model fused with random forest,'' \emph{IEEE Transactions on Image
  Processing}, vol.~24, no.~11, pp. 3282--3292, 2015.

\bibitem{9053634}
Z.~Tu, J.~Lin, Y.~Wang, B.~Adsumilli, and A.~C. Bovik, ``Bband index: A
  no-reference banding artifact predictor,'' in \emph{ICASSP 2020 - 2020 IEEE
  International Conference on Acoustics, Speech and Signal Processing
  (ICASSP)}, 2020, pp. 2712--2716.

\bibitem{chen2014modeling}
C.~Chen, L.~K. Choi, G.~De~Veciana, C.~Caramanis, R.~W. Heath, and A.~C. Bovik,
  ``Modeling the time—varying subjective quality of http video streams with
  rate adaptations,'' \emph{IEEE Trans. Image Process.}, vol.~23, no.~5, pp.
  2206--2221, 2014.

\bibitem{moorthymobile}
A.~K. Moorthy, L.~K. Choi, G.~de~Veciana, and A.~Bovik, ``Mobile video quality
  assessment database,'' in \emph{IEEE ICC Workshop on Realizing Advanced Video
  Optimized Wireless Networks}, 2012, pp. 7055--7059.

\bibitem{lin2015mcl}
J.~Y. Lin, R.~Song, C.-H. Wu, T.~Liu, H.~Wang, and C.-C.~J. Kuo, ``{MCL-V: A
  streaming }video quality assessment database,'' \emph{J. Vis. Commun. Image
  Represent.}, vol.~30, pp. 1--9, 2015.

\bibitem{keimel2010visual}
C.~Keimel, J.~Habigt, T.~Habigt, M.~Rothbucher, and K.~Diepold, ``Visual
  quality of current coding technologies at high definition {IPTV} bitrates,''
  presented at the IEEE MMSP, 2010, pp. 390--393.

\bibitem{mercat2020uvg}
A.~Mercat, M.~Viitanen, and J.~Vanne, ``Uvg dataset: 50/120fps {4K} sequences
  for video codec analysis and development,'' in \emph{ACM MMSys 2020}, pp.
  297--302.

\bibitem{cheon2017subjective}
M.~Cheon and J.-S. Lee, ``Subjective and objective quality assessment of
  compressed {4K UHD} videos for immersive experience,'' \emph{IEEE Trans.
  Circuits Syst. Video Technol}, vol.~28, no.~7, pp. 1467--1480, 2017.

\bibitem{li2016toward}
Z.~Li, A.~Aaron, I.~Katsavounidis, A.~Moorthy, and M.~Manohara, ``Toward a
  practical perceptual video quality metric,'' \emph{Netflix Tech Blog},
  vol.~6, p.~2, 2016.

\bibitem{video2000final}
\BIBentryALTinterwordspacing
{Video Quality Experts Group}, \emph{"Final report from the video quality
  experts group on the validation of objective quality metrics for video
  quality assessment"}, 2000 (accessed October 31,2020). [Online]. Available:
  \url{http://www. its. bldrdoc. gov/vqeg/projects/frtv phaseI}
\BIBentrySTDinterwordspacing

\bibitem{yodel2011consumer}
\BIBentryALTinterwordspacing
W.~Yodel, \emph{The consumer digital video library}, 2011. [Online]. Available:
  \url{https://cdvl.org/}
\BIBentrySTDinterwordspacing

\bibitem{song2013sjtu}
L.~Song, X.~Tang, W.~Zhang, X.~Yang, and P.~Xia, ``The {SJTU 4K} video sequence
  dataset,'' presented at the 5th QoMEX.\hskip 1em plus 0.5em minus 0.4em\relax
  IEEE, 2013, pp. 34--35.

\bibitem{union2012methodology}
{ITU}, ``Methodology for the subjective assessment of the quality of television
  pictures {ITU-R} recommendation {BT.} 500-13,'' Tech. Rep, Tech. Rep., 2012.

\bibitem{yu2019predicting}
X.~Yu, C.~G. Bampis, P.~Gupta, and A.~C. Bovik, ``Predicting the quality of
  images compressed after distortion in two steps,'' \emph{IEEE Trans. Image
  Process.}, vol.~28, no.~12, pp. 5757--5770, 2019.

\bibitem{mittal2012making}
A.~Mittal, R.~Soundararajan, and A.~C. Bovik, ``Making a “completely blind”
  image quality analyzer,'' \emph{IEEE Signal Process. Lett}, vol.~20, no.~3,
  pp. 209--212, 2012.

\bibitem{mittal2012no}
A.~Mittal, A.~K. Moorthy, and A.~C. Bovik, ``No-reference image quality
  assessment in the spatial domain,'' \emph{IEEE Trans. image Process.},
  vol.~21, no.~12, pp. 4695--4708, 2012.

\bibitem{kundu2017no}
D.~Kundu, D.~Ghadiyaram, A.~C. Bovik, and B.~L. Evans, ``No-reference quality
  assessment of tone-mapped hdr pictures,'' \emph{IEEE Trans. Image Process.},
  vol.~26, no.~6, pp. 2957--2971, 2017.

\bibitem{ye2012unsupervised}
P.~Ye, J.~Kumar, L.~Kang, and D.~Doermann, ``Unsupervised feature learning
  framework for no-reference image quality assessment,'' in \emph{CVPR
  2012}.\hskip 1em plus 0.5em minus 0.4em\relax IEEE, pp. 1098--1105.

\bibitem{korhonen2019two}
J.~Korhonen, ``Two-level approach for no-reference consumer video quality
  assessment,'' \emph{IEEE Trans. Image Process.}, vol.~28, no.~12, pp.
  5923--5938, 2019.

\bibitem{mittal2015completely}
A.~Mittal, M.~A. Saad, and A.~C. Bovik, ``A completely blind video integrity
  oracle,'' \emph{IEEE Trans. Image Process.}, vol.~25, no.~1, pp. 289--300,
  2015.

\bibitem{saad2012blind}
M.~A. Saad, A.~C. Bovik, and C.~Charrier, ``Blind image quality assessment: A
  natural scene statistics approach in the {DCT} domain,'' \emph{IEEE Trans.
  Image Process.}, vol.~21, no.~8, pp. 3339--3352, 2012.

\bibitem{ebenezer2020noreference}
\BIBentryALTinterwordspacing
J.~P. Ebenezer, Z.~Shang, Y.~Wu, H.~Wei, and A.~C. Bovik. (2020) No-reference
  video quality assessment using space-time chips. [Online]. Available:
  \url{https://arxiv.org/abs/2008.00031}
\BIBentrySTDinterwordspacing

\end{thebibliography}

\vspace{12pt}
\end{document}